\documentclass{aa}
\usepackage[varg]{txfonts}

\begin{document}

\title{Molecule mapping of HR8799b using OSIRIS on KECK:
  } 
\subtitle{Strong detection of water and carbon monoxide, but no methane}

\author{D. J. M. Petit dit de la Roche\inst{1}
  \and H. J. Hoeijmakers\inst{2}
     \and I. A. G. Snellen\inst{1}} 


\institute{Leiden Observatory, Leiden University, Postbus 9513, 2300RA Leiden, The Netherlands\\ e-mail: petit@strw.leidenuniv.nl \and Geneva Observatory, 51 chemin des Maillettes, 1290 Versoix, Switzerland}

\date{Received 7 may 2018 / Accepted 24 may 2018}

\abstract {In 2015, Barman et al. presented detections of absorption from water, carbon monoxide, and methane in the atmosphere of the directly imaged exoplanet HR8799b using integral field spectroscopy (IFS) with OSIRIS on the Keck II telescope. We recently devised a new method to analyse IFU data, called {\sl molecule mapping}, searching for high-frequency signatures of particular molecules in an IFU data cube.} {The aim of this paper is to use the molecule mapping technique to search for the previously detected spectral signatures in HR8799b using the same data, allowing a comparison of molecule mapping with previous methods.} {The medium-resolution H and K-band pipeline-reduced archival data were retrieved from the Keck archive facility. Telluric and stellar lines were removed from each spectrum in the data cube, after which the residuals were cross-correlated with model spectra of carbon monoxide, water and methane.} {Both carbon monoxide and water are clearly detected at high signal-to-noise, however, methane is not retrieved.}{Molecule mapping works very well on the OSIRIS data of exoplanet HR8799b. However, it is not evident why methane is detected in the original analysis, but not with the molecule mapping technique. Possible causes could be the presence of telluric residuals, different spectral filtering techniques, or the use of different methane models. We do note that in the original analysis methane was only detected in the K-band, while the H-band methane signal could be expected to be comparably strong. More sensitive observations with the JWST will be capable of confirming or disproving the presence of methane in this planet at high confidence.}


\maketitle

\section{Introduction}
The planetary system HR8799 is unique in that it hosts four directly-imaged planets \citep{Marois2008,Marois2010}. The A-star is estimated to have an age of 30 Myr \citep{Marois2010} and is located at a distance of 39 pc \citep{Marois2008}. The planets orbit at 15 to 70 au between a warm dust belt and a cold Kuiper belt-like structure \citep{Su2009, Booth2016}, implying orbital periods between 50 and 500 years. Indeed, recent time series observations clearly show the orbital motion of the planets \footnote{http://jasonwang.space/orbits.html}. HR8799b is the outermost planet, 1.7 arcseconds away from the host star, with a K-band magnitude of Ks=14.05, corresponding to a contrast of $\sim3300$ at these wavelengths \citep{Marois2008}. Comparisons of the planet's spectral energy distribution with planet evolutionary models provide a mass estimate of $M=5^{+2}_{-1} M_{\rm{J}}$ with a planet radius of $R=1.2\pm0.1 R_{\rm{J}}$ \citep{Currie2011,Soummer2011,Wright2011}.

Early observations with the integral field spectrograph OSIRIS \citep{OSIRIS2003,OSIRIS2006} on the Keck II telescope at $2.1 - 2.2\ \mu $m combined with flux measurements at other wavelengths show that the near-infrared spectrum of HR8799b is consistent with that of an early T-dwarf with an intermediate cloud deck and possibly a weak methane feature \citep{Bowler2010}. Atmospheric modeling by \citet{Madhusudhan2011} infers a thick cloud cover, and shows that the 3.3 $\mu $m flux of the planet is inconsistent with the expected fiducial methane abundance - requiring significantly less methane (see also \cite{Marley2012,Skemer2012}). Further observations by \citet{Oppenheimer2013} and \citet{Ingraham2014} also show no significant methane absorption. 

However, later observations with OSIRIS at H and K-band do provide evidence for methane  \citep{Barman2011,Barman2015}. A binned-down, low resolution spectral analysis of the IFU data \citep{Barman2011} shows strong water absorption in H and K, and weak methane and CO in K-band. Subsequent analysis at the full resolving power of R=4000 \citep{Barman2015} using cross-correlation techniques, seem to confirm the methane and carbon monoxide detections in K-band, but show no evidence for methane absorption in the H-band data, while this is expected from the absorption feature at 1.65 $\mu$m.  

In this paper, we apply a different analysis on the existing OSIRIS H and K-band data, also making use of cross-correlation techniques. The technique, dubbed {\it molecule mapping}, targets the high-frequency signatures of particular molecules in an IFU data cube, ultimately providing a measure of the amount of evidence for molecular absorption at all spatial locations in the IFU data cube \citep{Hoeijmakers2018}. The technique is described in detail in \cite{Hoeijmakers2018}, where it is applied to SINFONI/VLT data of the beta Pictoris system, showing strong absorption from both CO and H$_2$O at SNRs$\sim$15 at the location of exoplanet beta Pictoris b. First, a stellar reference spectrum is constructed using the spaxels near the star position. Subsequently, a scaled version of this stellar reference is subtracted from the spectra at each position in the IFU field of view. The redisual spectra are then cross-correlated with molecular templates. The cross-correlation co-adds the individual absorption lines of the planet emission spectrum constructively, while this is not the case for (residual) telluric and stellar features \citep{Hoeijmakers2018}. The method depends neither on field rotation, as is the case for Angular Differential Imaging (ADI; \cite{Marois2006}) methods, nor on wavelength scaling of the PSF, as for Spectral Differential Imaging (SDI; \cite{Mawet2012}) techniques, making it particularly powerful at small angular distances from the host star where residual speckles are strongest. Molecule mapping is different from the technique utilized by \cite{Barman2015}. In particular, they perform telluric calibrations using an early-type standard star, which implicitly also removes possible stellar absorption lines. 

The observations are described in Section \ref{sec:observations}, and the data analysis is  presented in Section \ref{sec:analysis}. The resulting molecule maps are shown and discussed in Section \ref{sec:results}.

\section{Observations}
\label{sec:observations}

The data of HR8799b were taken on four nights in July 2009 (H-band) and July 2013 (K-band; see Table  \ref{tab:exposures}) using the Natural guide star adaptive optics OH-Suppressing InfraRed Imaging Spectrograph (OSIRIS) on Keck II \citep{OSIRIS2006}, as part of programs U150O and U136O \citep{Barman2011,Barman2015}. OSIRIS utilizes a grid of lenslets to slice the field of view, after which a grating disperses the light from each lenslet resulting in a spectrum for each field-point. The data were taken in H-band mode and K-band mode, producing a nominal spectral resolution of R$\sim$4,000. Each data cube contains 16$\times$64 spectra at an angular scale of 0.02" per pixel. Time series of 6$\times$900s and 4$\times$900s in H-band, and two of 9$\times$600s in K-band were taken. 

\begin{table}[h]
	\centering
	\begin{tabular}{|c|c|c|c|}
		\hline
    	Band & Exposures & Exposure time & Date\\
    	\hline
   		H & 6 & 900s & 23-07-2009 \\
    	H & 4 & 900s & 30-07-2009\\
    	K & 9 & 600s & 25-07-2013\\
    	K & 9 & 600s & 26-07-2013\\
    	\hline 
    \end{tabular}
    \caption{\label{tab:exposures}Overview of the used OSIRIS observations.}
\end{table}

\section{Data analysis}
\label{sec:analysis}
\subsection{Background and telluric removal}
We used the pipeline-reduced datacubes retrieved from the Keck Archive, which provide the user with standardised and calibrated data cubes. These cubes were aligned and co-added such that the planet position was the same, resulting in one final data cube for each band. Two of the ten original H band cubes were excluded as the planet was not distinguishable from the background.

\begin{figure}
	\centering
    \resizebox{\hsize}{!}{\includegraphics{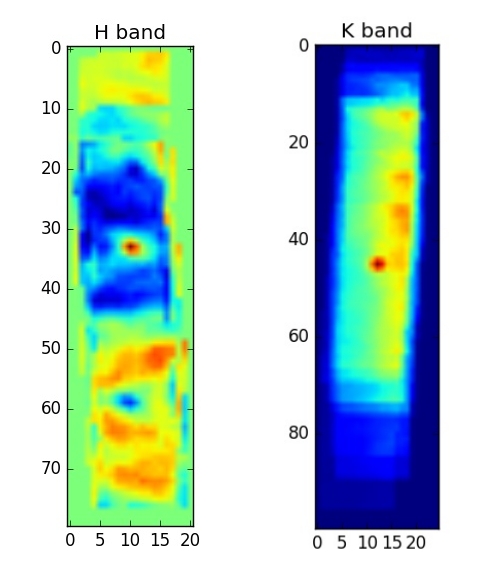}}
    \caption{\label{fig:whitelight}White light frames of the master data cubes in the H band and the K band.}
\end{figure}

The H band observations were performed in nodding mode with 0.5'' offsets perpendicular to the star direction, keeping the planet in view. This meant that no seperate background subtraction was necessary. However, in the K band there was significant background flux, meaning that the spectra contained a mix of planetary and stellar light with strong telluric lines. The measured spectrum in each pixel is decribed by,
\begin{equation}
	\label{eq:signal}
    \centering
	\begin{aligned}
		F_{\rm{x,y}}(\lambda) &= (A_{\rm{x,y}} * F_{\rm{s}}(\lambda)+B_{\rm{x,y}} * F_{\rm{p}}(\lambda)) \times S_{\rm{T}}(\lambda) \\
	\end{aligned}
\end{equation}
where $F_{\rm{s}}$ and $F_{\rm{p}}$ are the spectra of the star and planet respectively, and $S_{\rm{T}}$ is the telluric absorption spectrum. Constants $A_{\rm{x,y}}$ and $B_{\rm{x,y}}$ denote the relative contribution of the star and planet at each pixel location $x,y$. The latter is generally zero, except near or at the planet position. 

The first part of the data analysis was aimed at removing the contribution and contamination from the stellar and telluric spectrum from each spaxel. A Phoenix spectral model \citep{PHOENIXI}, with $T_{\rm{eff}}$ = 7200 K, log(g) = 4.50 cm sec$^{-2}$, solar metallicity, and convolved to the resolution of the observed spectra, was used as a template for the stellar spectrum. Subsequently, a reference spectrum was constructed by combining the spectra from a section of 25 by 10 pixels away from the planet position, which was divided by the stellar model to obtain the telluric spectrum. 
Each spaxel in the dataset is subsequently divided by the telluric spectrum to obtain a 'telluric free' data cube. As a final step, a stellar spectrum was weighted and subtracted from each spaxel such that the stellar lines were removed and only the planetary contribution remains. 

\subsection{Cross-correlation}

Because we focus our analysis on the high-frequency planet absorption features, each spaxel was high-pass filtered using a 6th-order Butterworth filter with a window of 10 nm in each band, corresponding to 50 and 40 pixels in the H and K bands respectively. This also removes any residual low-frequency stellar speckle contributions in the spectra. Data points that were more than 5$\sigma$ away from the mean were replaced by the local mean to prevent bad data points from influencing the cross correlation.

The data were cross-correlated with models from \cite{Models} of absorption spectra of methane, water, carbon monoxide and ammonia in the K-band, and of methane and water in the H-band, as carbon monoxide and ammonia have no strong features in this latter band. These models are the result of line-by-line model calculations using HITEMP \citep{rothman2010hitemp} and HITRAN 2008 \citep{rothman2009hitran} line databases, after which they were convolved to the resolution of the observed spectra and high-pass filtered as above. 
The cross-correlation was performed for each spaxel for radial velocities in the range from $-$1500 km/s to $+$1500km/s in steps of 35 km sec$^{-1}$.

\section{Results and Discussion}
\label{sec:results}
\subsection{Molecule maps}
\begin{figure}
	\centering
    \includegraphics[scale=.5]{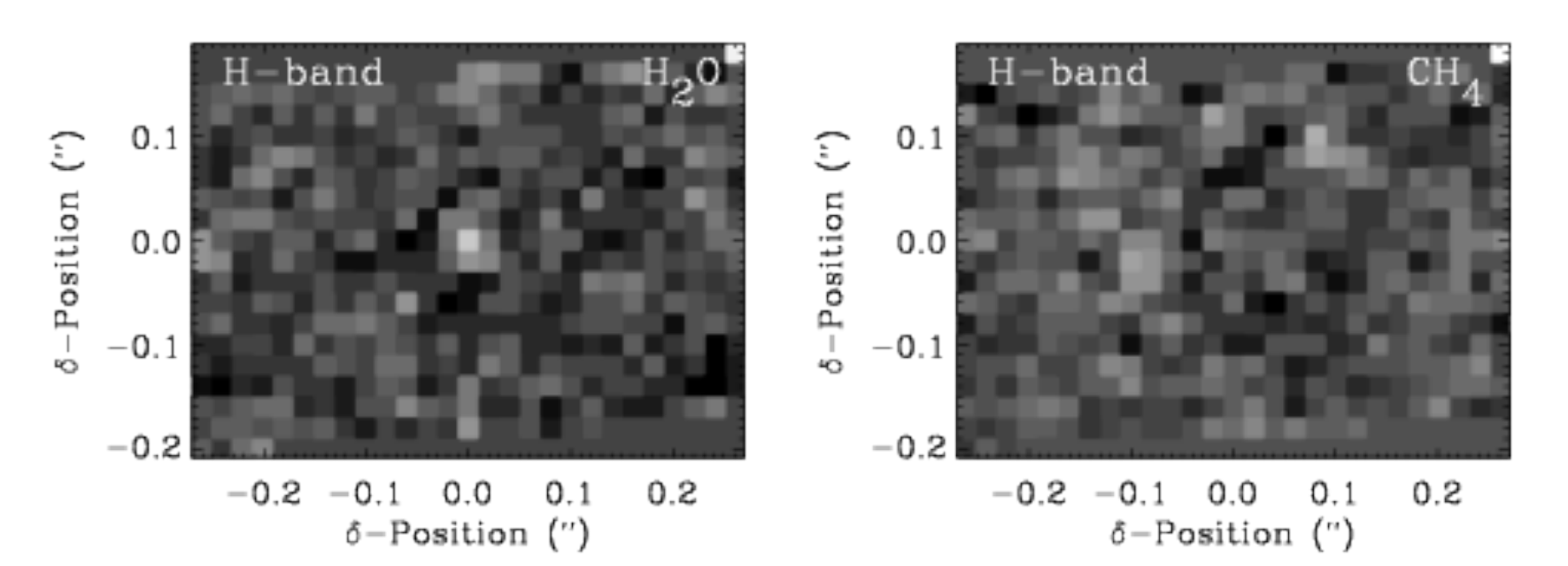}
    \caption{\label{fig:mmH}Molecule maps from the H-band data showing water on the left and methane on the right. The maps are centered on the location of the planet. The planet is detected in the map of water, but not in that of methane.}
\end{figure}

The resulting molecule maps consist of the cross-correlation signal from each spaxel at the planet radial velocity, here assumed to be 35 km sec$^{-1}$ (blueshifted) for all nights, originating from the combination of the barycentric velocity of the observatory (similar for all observations) and the system velocity of the star. The resolution of the cross-correlation function in both bands is also about 35 km sec$^{-1}$. Figure \ref{fig:mmH} presents the molecule maps from the H-band data, showing water and methane respectively. While water is marginally detected at SNR$\sim$2, there is no sign of methane. The SNRs are determined as the peak signal at the planet velocity divided by the standard devation of the cross-correlation function over a wide range of velocity ($\pm15000$ km sec$^{-1}$).

\begin{figure}
	\centering
    \includegraphics[scale=.5]{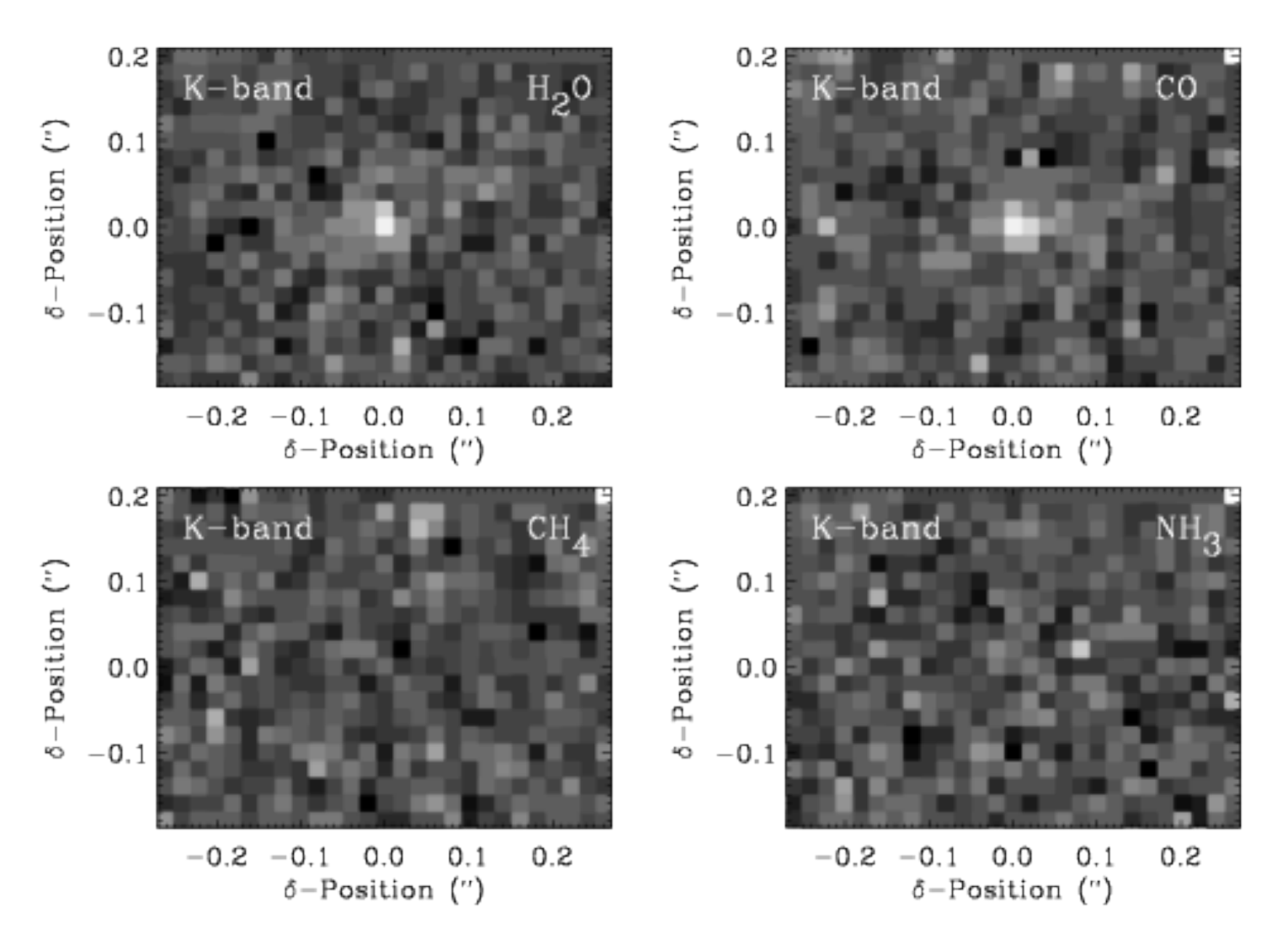}
    \caption{\label{fig:mmK} Molecule maps from the K-band data showing in the top row left to right carbon monoxide and water and in the bottom row methane and ammonia. The maps are centered on the location of the planet. The planet is detected in the maps of water and carbon monoxide, but not in the maps of methane and ammonia.}
\end{figure}

Figure \ref{fig:mmK} presents the molecule maps from the K-band data, showing from top-left to bottom-right water, carbon monoxide, methane, and ammonia. Both carbon monoxide and water are clearly detected at SNR$\sim$6, while again there is no sign of methane, and neither of ammonia. 

\begin{figure}
	\centering
    \resizebox{\hsize}{!}{\includegraphics{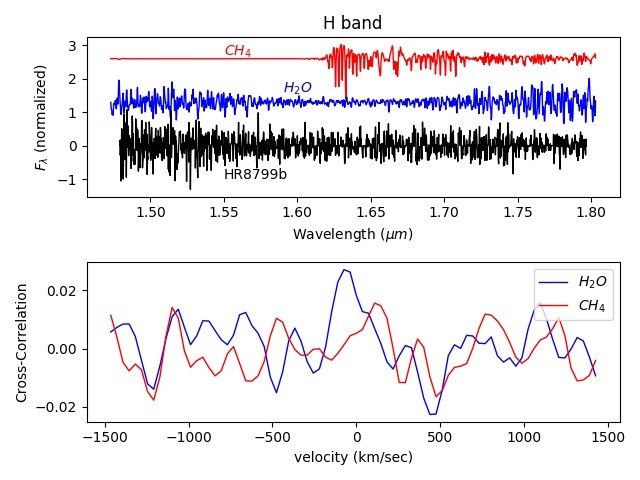}}
    \caption{\label{fig:cc-H-results} \emph{Top:} Filtered spectrum of HR8799b (black) and template spectra of water (blue) and methane (red) in the H band, offset by an arbitrary amount. \emph{Bottom:} Cross-correlation functions of the planet spectrum with the model spectra plotted above.}
\end{figure}

\begin{figure}
	\centering
    \resizebox{\hsize}{!}{\includegraphics{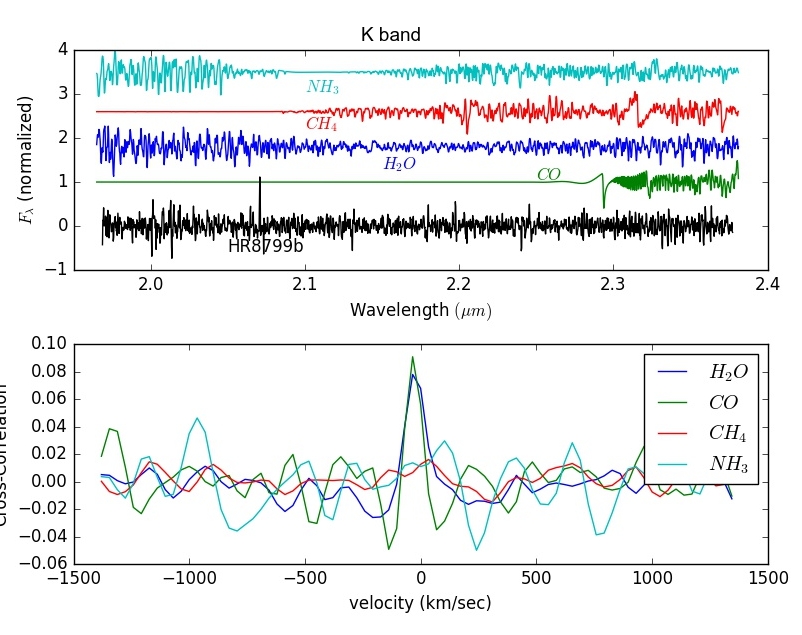}}
    \caption{\label{fig:cc-K-results} As Figure \ref{fig:cc-H-results}, but in the K band and with the addition of methane (green) and ammonia (cyan) model spectra and cross-correlation functions. The difference in the resolution of the CCF is due to the difference in resolution of the H and K band data.}
\end{figure}

The top panel of Figure \ref{fig:cc-H-results} shows the normalized and filtered water and methane model spectra used for the cross-correlation in H-band, and the processed spectrum at the planet position. The lower panel shows the cross-correlation functions for these molecules at this planet position, showing the low-confidence signal for water, and the absence of a methane signal. Figure \ref{fig:cc-K-results} shows the same for the K-band data, including carbon monoxide, water, methane and ammonia. The cross correlation functions show the clear detection for water and carbon monoxide at the planet velocity, but no sign of methane and ammonia. The resulting signal-to-noise ratios are summarized in Table \ref{tab:SNR}.

\begin{table}[h!]
	\centering
	\begin{tabular}{|c|c|c|}
		\hline
    	Molecule & H band & K band\\
    	\hline
   		$H_2O$ & 2.2 & 6.2 \\
    	$CH_4$ & 0.4 & 0.8 \\
    	$CO$ & - & 6.4 \\
    	$NH_3$ & - & 0.6 \\
    	\hline 
    \end{tabular}
    \caption{\label{tab:SNR}Signal to noise ratios of the cross-correlation functions in the H and K bands at a the planet radial velocity of -35 km sec$^{-1}$. }
\end{table}

No significant molecule signals are found away from the planet position. This means that the telluric line removal was successful, and that possible telluric residuals have not resulted in spurious signals.

\subsection{Comparison to earlier work}

We directly compare our results with those obtained by \cite{Barman2011} and \cite{Barman2015}, who used the same OSIRIS data - albeit we use a total of 50\% and 40\% fewer on-target exposures than the previous analysis due to the availability of reduced data cubes in the Keck archive. An equally powerful analysis would result in $\sim$30\% lower SNRs. Although  \cite{Barman2011} and \cite{Barman2015} do not explicitly quote a statistical significance for their detections, visual inspection of their cross-correlation functions indicate that our detections of H$_2$O and CO, in particular in the K-band, are significantly stronger. We also note that the velocity resolution of our cross-correlation functions are higher. This may in part explain why our analysis results in a higher confidence for these molecules.s

In contrast, while \cite{Barman2015} present a significant detection of methane in the K-band OSIRIS data, we see no sign of this molecule in our analysis. Possible causes could be the presence of telluric residuals in either of the analyses, the differences in spectral filtering techniques, or the use of different methane models. 
In particular, they perform telluric calibrations using an early-type standard star, while our analysis performs effectively a 'self-calibration' for telluric line removal using the data themselves. We also note that in the original analysis methane was only detected in the K-band, while the H-band methane signal could be expected to be comparably strong. Future more sensitive observations, e.g. with the JWST, should reveal whether methane is really present in the atmosphere of HR8799b or not.

\bibliographystyle{apa}
\bibliography{bibliography}

\end{document}